\documentstyle[12pt]{article}
\def\lsim{\raisebox{-4pt}{$\,\stackrel{\textstyle{<}}{\sim}\,$}}
\def\gsim{\raisebox{-4pt}{$\,\stackrel{\textstyle{>}}{\sim}\,$}}
\begin{document}
\begin{flushright}
\baselineskip=12pt
ACT-05/97\\
CTP-TAMU-10/97\\
DOE/ER/40717--40\\
\tt hep-ph/9702350
\end{flushright}

\begin{center}
\vglue 1.5cm
{\Large\bf New supersymmetric contributions to $t\to cV$}
\vglue 2.0cm
{\large Jorge L. Lopez$^1$, D.V. Nanopoulos$^{2,3}$, and Raghavan
Rangarajan$^3$}
\vglue 1cm
\begin{flushleft}
$^1$ Bonner Nuclear Lab, Department of Physics, Rice University\\ 6100 Main
Street, Houston, TX 77005, USA\\
$^2$Center for Theoretical Physics, Department of Physics, Texas A\&M
University\\ College Station, TX 77843--4242, USA\\
$^3$Astroparticle Physics Group, Houston Advanced Research Center (HARC)\\
The Mitchell Campus, The Woodlands, TX 77381, USA\\
\end{flushleft}
\end{center}

\vglue 1.5cm
\begin{abstract}
We calculate the electroweak-like one-loop supersymmetric contributions to the
rare and flavor-violating decay of the top quark into a charm quark and a gauge
boson: $t\to c V$, with $V=\gamma,Z,g$. We consider loops of both charginos
and down-like squarks (where we identify and correct an error in the
literature) and neutralinos and up-like squarks (which have not been calculated
before). We also account for left-right and generational squark mixing. Our
numerical results indicate that supersymmetric contributions to $t\to cV$ can 
be upto 5 orders of magnitude larger than their Standard Model counterparts.
However, they still fall short of the sensitivity expected at the 
next-generation top-quark factories.
\end{abstract}

\vspace{0.5cm}
\begin{flushleft}
{\tt lopez@physics.rice.edu\\
dimitri@phys.tamu.edu\\
raghu@diana.tdl.harc.edu\\}
\smallskip
February 1997\\
\end{flushleft}
\newpage
\setcounter{page}{1}
\pagestyle{plain}
\baselineskip=14pt

\section{Introduction}
The discovery of the top quark by the CDF and D0 Collaborations \cite{CDFD0}
at Fermilab, and its subsequent mass determination ($m_t=175\pm6\,{\rm GeV}$)
have initiated a new era in particle spectroscopy. However, unlike the lighter
quarks, the top quark is not expected to form any bound states, and therefore
its mass and decay branching ratios may be determined more precisely, both
theoretically and experimentally. Upcoming (Run II -- Main Injector) and
proposed (Run III -- TeV33) runs at the Tevatron will yield large numbers of
top quarks, as will be the case at the LHC, turning these machines effectively
into `top factories.' Even though higher precision in the determination of
$m_t$ is expected (it is already known to 3\%), more valuable information
should come from the precise determination of its branching fractions into
tree-level and rare (and perhaps even `forbidden') decay channels.

The purpose of this paper is to study one class of such rare decay modes:
$t\to cV$, with $V=\gamma,Z,g$. The particular case of $t\to cg$ has received
some phenomenological attention recently as a means to probe the scale at
which such new and unspecified interactions might turn on \cite{tcg}. Our
purpose here is to consider an explicit realization of this coupling within
the framework of low-energy supersymmetry. This is different in spirit from
the line of work in Ref.~\cite{tcg}, as the effective mass scale at which such
vertices `turn on' is determined here by the interactions of presumably rather
light sparticles. Within supersymmetry, the $t\to cV$ vertex was first
contemplated in Ref.~\cite{LOY}, where the one-loop QCD-like (loops of gluinos
and squarks) and electroweak-like (loops of charginos and down-like squarks)
contributions were calculated. The QCD-like supersymmetric corrections were
subsequently re-evaluated and generalized in Ref.~\cite{CHK}, which pointed
out an inconsistency in the corresponding results of Ref.~\cite{LOY}. Here
we study the electroweak-like supersymmetric contributions to $t\to cV$.
We reconsider the chargino--down-like-squark loops and point out and correct an
inconsistency, essentially a lack of gauge invariance because of the apparent
omission of a term, in the corresponding results of Ref.~\cite{LOY}. We also
consider for the first time the neutralino--up-like-squark loops, and include
the effects of left-right and generational squark mixing.

Our numerical results indicate that for typical values of the parameters one
gets a large enhancement over Standard Model predictions of top-quark decays to
gauge bosons \cite{tcV-SM}.  
For the most optimistic values of the parameters the enhancement can be as large
as five orders of magnitude.
However, even for the most optimistic values of
the parameters, such rare decay channels fall short of the expected
sensitivity of the next-generation top factories. 

\section{Analytical Results}
\label{sec:formulas}
In this section we obtain the one-loop electroweak-like supersymmetric
effective top-quark-charm-quark-gauge-boson vertex by considering loops
involving charginos and neutralinos, including the effects from left-right
and generational squark mixing. We then present the decay rates of the
top-quark to the charm-quark and a gauge boson.

\begin{figure}[p]
\vspace{6in}
\includegraphics{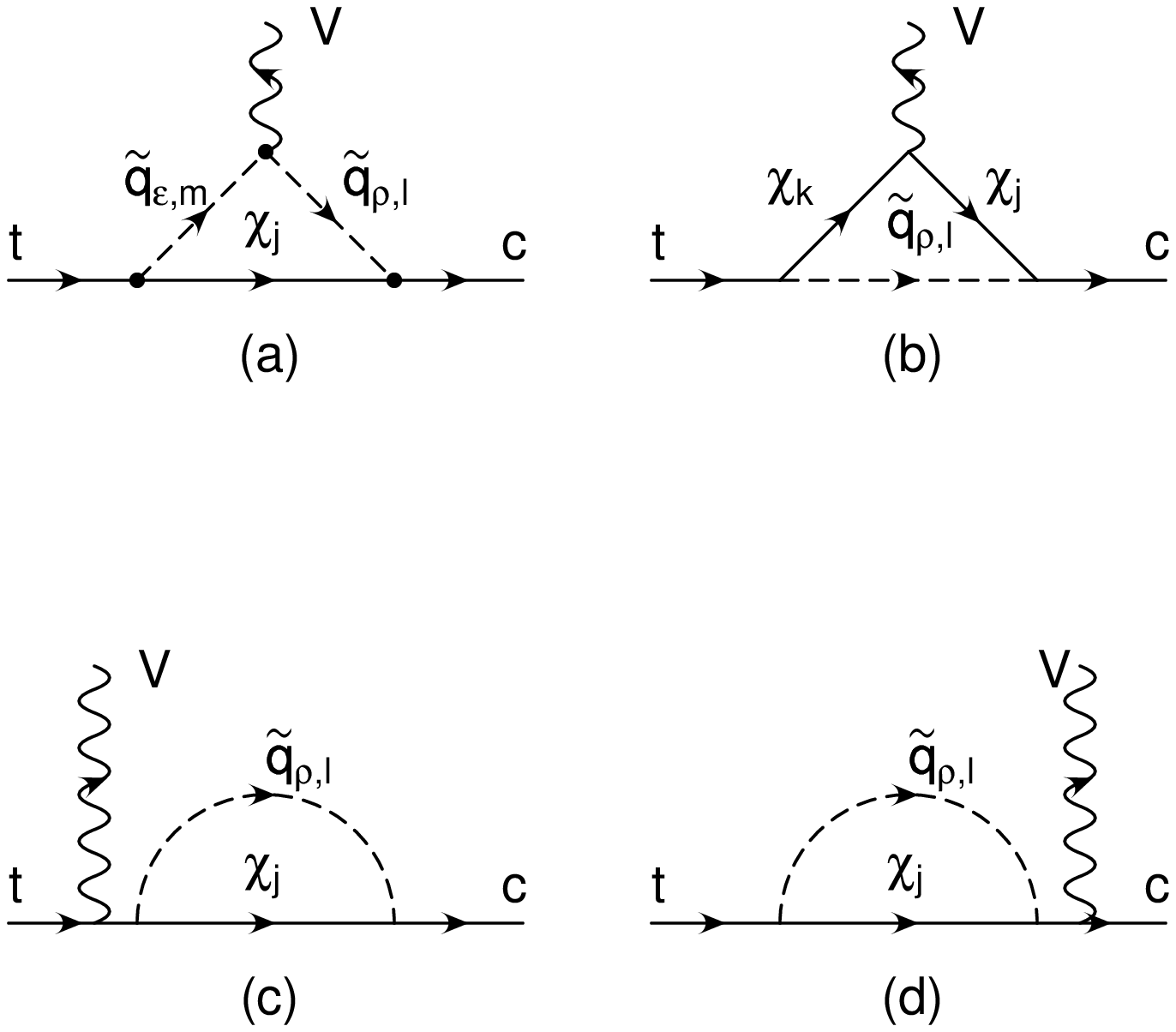}
\vspace{1cm}
\caption{Feynman diagrams for one-loop electroweak-like supersymmetric
contributions to the $t$-$c$-$V$ ($V=\gamma,Z,g$) vertex. In the figure $\chi$
represents the chargino or the neutralino and $\tilde q$ represents the
down-type or up-type squarks respectively. The subscripts are explained in the
text. The arrows on the squark lines indicate the direction of flow of
flavor; the arrows on the gauge bosons indicate the direction of momentum flow.
Diagram (b) is absent for $V=g$, and for $V=\gamma$ when $\chi=\chi^0$.}
\label{fig:diagrams}
\end{figure}

The invariant amplitude for top-quark decay to a charm-quark and a gauge
boson can be written as
\begin{equation}
M=M_0+ \delta M
\label{eq:A1}
\end{equation}
where $M_0$ is the tree-level amplitude and $\delta M$ is the first-order
supersymmetric correction.  As there are no explicit flavor-violating $tcV$
couplings in the Lagrangian $M_0=0$, whereas $\delta M$ is given by
\begin{equation}
i \delta M=\bar u(p_2)\, V^\mu  u(p_1)\, \epsilon_\mu(k,\lambda)
\,,
\label{eq:deltaM}
\end{equation}
where $p_1, p_2$, and $k$ are the momenta of the incoming top-quark,
outgoing charm-quark, and outgoing gauge boson respectively, and
$\epsilon_\mu(k,\lambda)$ is the polarization vector for the outgoing gauge
boson. The vertices $V^\mu$ may be written as
\begin{eqnarray}
V^\mu(tcZ)&=&-i \gamma^\mu (P_L F_{Z1}+ P_R F_{Z1}^\prime)
+ k_\nu\sigma^{\mu\nu}(P_R F_{Z2} + P_L F_{Z2}^\prime)\, ,
\label{eq:VtcZ}\\
V^\mu(tc\gamma)&=&-i \gamma^\mu (P_L F_{\gamma1}+ P_R F_{\gamma1}^\prime)
+ k_\nu\sigma^{\mu\nu}(P_R F_{\gamma2} + P_L F_{\gamma2}^\prime)\, ,
\label{eq:Vtcgamma}\\
V^\mu(tcg)&=&-i T^a\gamma^\mu (P_L F_{g1}+ P_R F_{g1}^\prime)
+  T^a k_\nu\sigma^{\mu\nu}(P_R F_{g2} + P_L F_{g2}^\prime)\, .
\label{eq:Vtcg}
\end{eqnarray}
where as usual we have defined $P_{R,L}={1\over2}(1\pm\gamma_5)$ and
$\sigma^{\mu\nu}={i\over 2}[\gamma^\mu,\gamma^\nu]$.  $T^a$ are the generators
of $SU(3)_C$.  The form
factors $F_{1,2}$ and $F_{1,2}^\prime$ encode the loop functions and depend on
the various masses in the theory. The Feynman rules used to obtain them are
given in Refs.~\cite{GH,BBMR} and the corresponding Feynman diagrams are shown
in Fig.~\ref{fig:diagrams}.  The vertices are derived assuming $p_1-p_2-k=0$.

The form factors for the electroweak-like corrections due to loops involving
charginos and strange- and bottom-squarks are given by
\begin{eqnarray}
F_{i1}^c&=&{1\over 4\pi^2}\sum_{j=1}^2\sum_{\rho=1}^2\sum_{l=1}^2
\sum_{\epsilon=1}^2\sum_{m=1}^2 \biggl\{
A_c^{j,\rho,l}D_c^{j,\epsilon,m} E_{ic}^{\rho,\epsilon}\,
[m_t^2(c_{12}+c_{23}-c_{11}-c_{21}) -2c_{24}\nonumber\\
&&+m_c^2(c_{23}-c_{12})]
+ B_c^{j,\rho,l} C_c^{j,\epsilon,m}E_{ic}^{\rho,\epsilon}
 m_c m_t\,(2c_{23}-c_{11}-c_{21})\nonumber\\
&&+A_c^{j,\rho,l}C_c^{j,\epsilon,m} E_{ic}^{\rho,\epsilon}
 m_t m_{\chi^\pm_j}  (c_0+c_{11})\nonumber\\
&&+ B_c^{j,\rho,l} D_c^{j,\epsilon,m}E_{ic}^{\rho,\epsilon}
 m_c  m_{\chi^\pm_j}(c_0+c_{11})
\biggr\}^{(-p_1,k,\chi^\pm_j,\tilde \epsilon_m,\tilde \rho_l)}\nonumber\\
&&+{1\over 2\pi^2}\sum_{j=1}^2\sum_{k=1}^2\sum_{\rho=1}^2 \sum_{l=1}^2\biggl\{
B_c^{j,\rho,l} C_c^{k,\rho,l} F_{ic}^{j,k}
m_c m_t\, (c_{21}+c_{22}-2c_{23})\nonumber\\
&&+A_c^{j,\rho,l}D_c^{k,\rho,l} E_{ic}^{\prime j,k}\,
 [m_c^2 c_{22}-m_i^2(c_{12}+c_{23})-2c_{24}+{\textstyle{1\over2}}]
\nonumber\\
&&+B_c^{j,\rho,l} D_c^{k,\rho,l} F_{ic}^{j,k} m_c  m_{\chi^\pm_k}
(c_{12}-c_{11})
\nonumber\\
&&
+A_c^{j,\rho,l} C_c^{k,\rho,l} F_{ic}^{j,k} m_t  m_{\chi^\pm_j} (c_{12}-c_{11})
+A_c^{j,\rho,l}D_c^{k,\rho,l} F_{ic}^{j,k} m_{\chi^\pm_j}m_{\chi^\pm_k} c_0
\biggr\}^{(-p_1,p_2,\chi^\pm_k,\tilde \rho_l,\chi^\pm_j)}\nonumber\\
&&+{1\over 2\pi^2}{1\over m_t^2-m_c^2}
\sum_{j=1}^2\sum_{\rho=1}^2 \sum_{l=1}^2\biggl\{
A_c^{j,\rho,l}D_c^{j,\rho,l} H_{ic} m_t^2 (-B1)
+A_c^{j,\rho,l}C_c^{j,\rho,l} H_{ic} m_t m_{\chi^\pm_j} B0\nonumber\\
&&+B_c^{j,\rho,l}C_c^{j,\rho,l} H_{ic} m_c m_t (-B1)
+B_c^{j,\rho,l}D_c^{j,\rho,l} H_{ic} m_c  m_{\chi^\pm_j} B0
\biggr\}^{(-p_1,\chi^\pm_j,\tilde \rho_l)}\nonumber\\
&&+{1\over 2\pi^2}{1\over m_t^2-m_c^2}
\sum_{j=1}^2\sum_{\rho=1}^2 \sum_{l=1}^2\biggl\{
A_c^{j,\rho,l}D_c^{j,\rho,l} H_{ic} m_c^2B1
+A_c^{j,\rho,l}C_c^{j,\rho,l} H_{ic} m_t m_{\chi^\pm_j} (-B0)\nonumber\\
&&+B_c^{j,\rho,l}C_c^{j,\rho,l} H_{ic} m_c m_t B1
+B_c^{j,\rho,l}D_c^{j,\rho,l} H_{ic} m_c  m_{\chi^\pm_j} (-B0)
\biggr\}^{(-p_2,\chi^\pm_j,\tilde \rho_l)}\, ,
\label{eq:Fi1c}\\
F_{i1}^{\prime c}&=&F_{i1}^c(A,B,C,D,E^\prime,F,H
\rightarrow B,A,D,C,F,E^\prime,G)\, ,
\label{eq:Fi1cprime}\\
F_{i2}^c&=&{1\over 4\pi^2}\sum_{j=1}^2\sum_{\rho=1}^2\sum_{l=1}^2
\sum_{\epsilon=1}^2\sum_{m=1}^2 \biggl\{
A_c^{j,\rho,l} D_c^{j,\epsilon,m} E_{ic}^{\rho,\epsilon}
 m_t(c_{12}+c_{23}-c_{11}-c_{21})
\nonumber\\
&&+A_c^{j,\rho,l} C_c^{j,\epsilon,m} E_{ic}^{\rho,\epsilon}
 m_{\chi^\pm_j} (c_0+c_{11})
+B_c^{j,\rho,l} C_c^{j,\epsilon,m} E_{ic}^{\rho,\epsilon} m_c (c_{23}-c_{12})
\biggr\}^{(-p_1,k,\chi^\pm_j,\tilde \epsilon_m,\tilde \rho_l)}\nonumber\\
&&+{1\over 2\pi^2}\sum_{j=1}^2\sum_{k=1}^2\sum_{\rho=1}^2 \sum_{l=1}^2\biggl\{
A_c^{j,\rho,l}D_c^{k,\rho,l} E_{ic}^{\prime j,k}
m_t(c_{11}+c_{21}-c_{12}-c_{23})
\nonumber\\
&&+B_c^{j,\rho,l} C_c^{k,\rho,l} F_{ic}^{j,k} m_c  (c_{22}-c_{23})
+A_c^{j,\rho,l} C_c^{k,\rho,l} E_{ic}^{\prime j,k}  m_{\chi^\pm_k}
(-c_{11}-c_0)
\nonumber\\
&&+A_c^{j,\rho,l}C_c^{k,\rho,l} F_{ic}^{j,k} m_{\chi^\pm_j} c_{12}
\biggr\}^{(-p_1,p_2,\chi^\pm_k,\tilde \rho_l,\chi^\pm_j)}\, ,
\label{eq:Fi2c}\\
F_{i2}^{\prime c}&=&F_{i2}^c(A,B,C,D,E^\prime,F
\rightarrow B,A,D,C,F,E^\prime)\, ,
\label{eq:Fi2cprime}
\end{eqnarray}
In these expressions $i=Z,\gamma,g$, the sums over $j,k=1,2$ run over the two
chargino mass eigenstates, $\rho,\epsilon=2,3$ represent strange- and
bottom-squarks (we ignore the mixing with the down-squark), and $l,m=1,2$
represent a sum over squark mass eigenstates which are obtained from the
$\tilde q_{L,R}$ gauge eigenstates via: $\tilde q_{\rho 1}=\cos\theta_\rho
\tilde q_{\rho L} +\sin\theta_\rho \tilde q_{\rho R}$ and $\tilde q_{\rho 2}
=-\sin\theta_\rho \tilde q_{\rho L} +\cos\theta_\rho \tilde q_{\rho R}$.
The various $B$ and $c$ functions in the above expressions are the well
documented Passarino-Veltman functions \cite{PV} (adapted to our metric where
$p^2_i=m^2_i$); the arguments of the $B$ and $c$ functions are indicated by the
superscripts on the braces in Eqs.~(\ref{eq:Fi1c})--(\ref{eq:Fi2cprime}).  For
example, the arguments of the $c$ functions that appear within the first brace
of Eq.~(\ref{eq:Fi1c}) are $(-p_1,k,m_{\chi^\pm_k},m_{\tilde q_{\epsilon m}},
m_{\tilde q_{\rho l}})$ while the arguments of the $B$ functions that appear
in the third brace of Eq.~(\ref{eq:Fi1c}) are $(-p_1, m_{\chi^\pm_k},
m_{\tilde q_{\rho l}})$. (Note that the Passarino-Veltman functions depend only
on the square of their arguments.) The Passarino-Veltman functions contain
infinities which cancel each other out, as they should since there is no
$t$-$c$-$V$ vertex in the Lagrangian. The coefficient functions are given by
\begin{eqnarray}
A_c^{j,\rho,l}&=&{g\over2}
\biggl[
-U_{j1}(K\Gamma_2^\dagger)_{2\rho}
{\cos\theta_\rho \; (l=1)\brace-\sin\theta_\rho \; (l=2)}
\nonumber\\
&&\quad\quad+{1\over \sqrt 2 m_W \cos\beta} U_{j2} (K M_d B_2^\dagger)_{2\rho}
{\sin\theta_\rho \;(l=1)\brace \cos\theta_\rho\;(l=2)}
\biggr]\, ,
\label{eq:Ac}\\
B_c^{j,\rho,l}&=&{g\over2}\left[ {m_c\over \sqrt 2 m_W \sin\beta} V_{j2}^*
(K\Gamma_2^\dagger)_{2\rho}
{\cos\theta_\rho \; (l=1)\brace-\sin\theta_\rho \; (l=2)}
\right]\, ,
\label{eq:Bc}\\
C_c^{j,\rho,l}&=&{g\over2}\left[ {m_t\over \sqrt 2 m_W \sin\beta} V_{j2}
(\Gamma_2K^\dagger)_{\rho3}
{\cos\theta_\rho \; (l=1)\brace-\sin\theta_\rho \; (l=2)}
\right]\, ,
\label{eq:Cc}\\
D_c^{j,\rho,l}&=&{g\over2}
\biggl[
-U_{j1}^*(\Gamma_2K^\dagger)_{\rho3}
{\cos\theta_\rho \; (l=1)\brace-\sin\theta_\rho \; (l=2)}
\nonumber\\
&&\quad\quad
+{1\over \sqrt 2 m_W \cos\beta} U_{j2}^* (B_2 M_d K^\dagger)_{\rho3}
{\sin\theta_\rho \; (l=1)\brace \cos\theta_\rho\; (l=2)}
\biggr]\, ,
\label{eq:Dc}\\
E_{Zc}^{\rho,\epsilon}&=&-{g\over\cos\theta_W}\left[
-{1\over2}\sum_{p=2}^3\Gamma_{QL}^{\rho p}\Gamma_{QL}^{*\epsilon p}
+{1\over3}\sin^2\theta_W\delta^{\rho\epsilon} \right] \, ,
\label{eq:EZc}\\
E_{\gamma c}^{\rho,\epsilon}&=&{e\over3}\delta^{\rho\epsilon}\, ,
\label{eq:Egammac}\\
E_{gc}^{\rho,\epsilon}&=&-g_s\delta^{\rho\epsilon}\, ,
\label{eq:Egc}\\
E_{Zc}^{\prime j,k}&=&{g\over2\cos\theta_W}\left[-U_{j1}^*U_{k1}
-{1\over2}U_{j2}^*U_{k2} +\sin^2\theta_W\delta^{jk} \right] \, ,
\label{eq:EprimeZc}\\
E_{\gamma c}^{\prime j,k}&=&-{e\over2}\delta^{jk}\, ,
\label{eq:Eprimegammac}\\
E_{g c}^{\prime j,k}&=&0 \, ,
\label{eq:Eprimegc}\\
F_{Zc}^{j,k}&=&{g\over2\cos\theta_W}\left[-V_{j1}V_{k1}^*
-{1\over2}V_{j2}V_{k2}^* +\sin^2\theta_W\delta^{jk} \right] \, ,
\label{eq:FZc}\\
F_{\gamma c}^{j,k}&=&-{e\over2}\delta^{jk}\, ,
\label{eq:Fgammac}\\
F_{g c}^{j,k}&=&0 \, ,
\label{eq:Fgc}\\
G_{Zc}&=&-{g\over2\cos\theta_W}\left(-{2\over3}\sin^2\theta_W\right)\, ,
\label{eq:GZc}\\
G_{\gamma c}&=&-{e\over3}\, ,
\label{eq:Ggammac}\\
G_{gc}&=&-{g_s\over2}\, ,
\label{eq:Ggc}\\
H_{Zc}&=&-{g\over2\cos\theta_W}\left({1\over2}-{2\over3}\sin^2\theta_W\right)\,
,
\label{eq:HZc}\\
H_{\gamma c}&=&-{e\over3}\, ,
\label{eq:Hgammac}\\
H_{gc}&=&-{g_s\over2}\, ,
\label{eq:Hgc}\, .
\end{eqnarray}
The chargino mixing matrices $U_{ij}$ and $V_{ij}$ and the generational mixing
matrices $K,\Gamma_2$ and $B_2$ which appear in these expressions are defined
in Ref.~\cite{GH}, $M_d$ is the diagonal matrix $(m_d, m_s, m_b)$ and
$\Gamma_{QL}$ is the squark mixing matrix defined in Ref.~\cite{BBMR}.

In deriving the above form factors, we have used the following relations:
\begin{eqnarray}
\sum_{\lambda} k^\mu k^\nu \epsilon_\mu(k,\lambda) \epsilon_\nu(k,\lambda)&=&0
\, ,
\label{eq:id1}\\
\bar u(p_2) p_1^\mu P_{R,L} u(p_1)&=&
\bar u(p_2)[m_t\gamma^\mu P_{L,R} +i p_{1 \nu} \sigma^{\mu\nu} P_{R,L}]
u(p_1) \, ,
\label{eq:id2}\\
\bar u(p_2) p_2^\mu P_{R,L} u(p_1)&=&
\bar u(p_2)[m_c\gamma^\mu P_{R,L} -i p_{2 \nu} \sigma^{\mu\nu} P_{R,L}]
u(p_1) \, ,
\label{eq:id3}\\
\bar u(p_2) (p_1+p_2)^\mu P_{R,L} u(p_1)\epsilon_\mu(k,\lambda)
&=&\bar u(p_2) 2p_1^\mu P_{R,L} u(p_1) \epsilon_\mu(k,\lambda)\nonumber\\
&=&\bar u(p_2) 2p_2^\mu P_{R,L} u(p_1)\epsilon_\mu(k,\lambda)\nonumber\\
&=&\bar u(p_2)[m_t\gamma^\mu P_{L,R} +m_c\gamma^\mu P_{R,L}\nonumber\\
&&\quad\qquad +i k_{\nu} \sigma^{\mu\nu} P_{R,L}]
u(p_1)\epsilon_\mu(k,\lambda) \,.
\label{eq:id4}
\end{eqnarray}
(The first two equalities in Eq.~(\ref{eq:id4}) are only true when one takes
the absolute square of both sides of the equation and uses Eq.~(\ref{eq:id1}).)

Our results above for the chargino-squark loops disagree with those of
Ref.~\cite{LOY} in the limit of $m_c=0$. We have an additional term
\begin{equation}
{1\over 2\pi^2}\sum_{j=1}^2\sum_{\rho=1}^2\sum_{l=1}^2
\sum_{\epsilon=1}^2\sum_{m=1}^2 \biggl\{
A_c^{j,\rho,l}D_c^{k,\rho,l} F_{ic}^{j,k} m_{\chi^\pm_j}m_{\chi^\pm_k} c_0
\biggr\}^{(-p_1,p_2,\chi^\pm_k,\tilde \rho_l,\chi^\pm_j)}\nonumber
\end{equation}
in $F_{i1}^c$.  This term is required by gauge invariance, {\em i.e.}, it is
needed to ensure that the coefficient of $\gamma^\mu$ in
Eqs.~(\ref{eq:Vtcgamma}) and and (\ref{eq:Vtcg}) vanishes
\cite{peskinschroeder} for the massless gauge bosons.

The form factors for the electroweak-like corrections due to loops involving
neutralinos and top- and charm-squarks are given by
Eqs.~(\ref{eq:Fi1c})--(\ref{eq:Fi2cprime}) with $m_{\chi^\pm}$ replaced by
$m_{\chi^0}$ and with the coefficients $A_c$--$F_c$ replaced by
\begin{eqnarray}
A_n^{j,\rho,l}&=&-{1\over\sqrt2}\left[
{2\over3}eN_{j1}^\prime+{g\over\cos\theta_W}({1\over2}-{2\over3}\sin^2\theta_W)
N_{j2}^\prime\right](\Gamma_1^\dagger)_{2\rho}
{\cos\theta_\rho \; (l=1) \brace -\sin\theta_\rho \; (l=2)}
\nonumber\\
&&-{1\over\sqrt2}\left[
{g\over2m_W\sin\beta}N_{j4}\right](M_uB_1^\dagger)_{2\rho}
{\sin\theta_\rho \;(l=1)\brace \cos\theta_\rho \;(l=2)}\, ,
\label{eq:An}\\
B_n^{j,\rho,l}&=&-{1\over\sqrt2}\left[
{g\over2m_W\sin\beta}N_{j4}^*\right](M_u\Gamma_1^\dagger)_{2\rho}
{\cos\theta_\rho \;(l=1)\brace -\sin\theta_\rho\;(l=2)}
\nonumber\\
&&+{1\over\sqrt2}\left[
{2\over3}eN_{j1}^{\prime*}-{g\over\cos\theta_W}({2\over3}\sin^2\theta_W)
N_{j2}^{\prime*}\right](B_1^\dagger)_{2\rho}
{\sin\theta_\rho \;(l=1)\brace \cos\theta_\rho\;(l=2)}\, ,
\label{eq:Bn}\\
C_n^{j,\rho,l}&=&-{1\over\sqrt2}\left[
{g\over2m_W\sin\beta}N_{j4}\right](\Gamma_1M_u)_{\rho3}
{\cos\theta_\rho \;(l=1)\brace -\sin\theta_\rho\;(l=2)}
\nonumber\\
&&+{1\over\sqrt2}\left[
{2\over3}eN_{j1}^{\prime}-{g\over\cos\theta_W}({2\over3}\sin^2\theta_W)
N_{j2}^{\prime}\right](B_1)_{\rho3}
{\sin\theta_\rho\;(l=1) \brace \cos\theta_\rho\;(l=2)}\, ,
\label{eq:Cn}\\
D_n^{j,\rho,l}&=&-{1\over\sqrt2}\left[
{2\over3}eN_{j1}^{\prime*}+{g\over\cos\theta_W}
({1\over2}-{2\over3}\sin^2\theta_W)
N_{j2}^{\prime*}\right](\Gamma_1)_{\rho3}
{\cos\theta_\rho \;(l=1)\brace -\sin\theta_\rho\;(l=2)}
\nonumber\\
&&-{1\over\sqrt2}\left[
{g\over2m_W\sin\beta}N_{j4}^*\right](B_1M_u)_{\rho3}
{\sin\theta_\rho \;(l=1)\brace \cos\theta_\rho\;(l=2)}\, ,
\label{eq:Dn}\\
E_{Zn}^{\rho,\epsilon}&=&-{g\over\cos\theta_W}\left[
{1\over2}\sum_{p=2}^3\Gamma_{QL}^{\rho p}\Gamma_{QL}^{*\epsilon p}
-{2\over3}\sin^2\theta_W\delta^{\rho\epsilon} \right] \, ,
\label{eq:EZn}\\
E_{\gamma n}^{\rho,\epsilon}&=&-{2\over3}e\delta^{\rho\epsilon}\, ,
\label{eq:Egamman}\\
E_{gn}^{\rho,\epsilon}&=&-g_s\delta^{\rho\epsilon}\, ,
\label{eq:Egn}\\
E_{Zn}^{\prime j,k}&=&{g\over2\cos\theta_W}\left[
{1\over2}N_{j3}^{\prime*}N_{k3}^\prime -{1\over2}N_{j4}^{\prime*}N_{k4}^\prime
 \right] \, ,
\label{eq:EprimeZn}\\
E_{\gamma n}^{\prime j,k}&=&0\, ,
\label{eq:Eprimegamman}\\
E_{g n}^{\prime j,k}&=&0 \, ,
\label{eq:Eprimegn}\\
F_{Zn}^{j,k}&=&{g\over2\cos\theta_W}\left[
-{1\over2}N_{j3}^{\prime}N_{k3}^{\prime*} +{1\over2}N_{j4}^{\prime}
N_{k4}^{\prime*}\right]\, ,
\label{eq:FZn}\\
F_{\gamma n}^{j,k}&=&0 \, ,
\label{eq:Fgamman}\\
F_{g n}^{j,k}&=&0 \,.
\label{eq:Fgn}
\end{eqnarray}
The neutralino mixing matrices $N_{ij}$ and $N_{ij}^\prime$ and the
generational mixing matrices $\Gamma_1$ and $B_1$, are defined in
Ref.~\cite{GH}, and $M_u$ is the diagonal matrix $(m_u, m_c, m_t)$.
In Eqs.~(\ref{eq:Fi1c})--(\ref{eq:Fi2cprime}) the sums over $j,k$ now run from
1--4 over the four neutralino mass eigenstates, and $\rho,\epsilon,p=2,3$
represent charm- and top-squarks (we ignore the mixing with the up-squark);
$l,m=1,2$ represent a sum over squark mass eigenstates, as earlier.
The coefficients $G$ and $H$ are unaltered.

While we have included the charm-quark mass in the form factors above for
completeness, we set $m_c=0$ hereafter. The supersymmetric electroweak-like
contribution to the decay rates is then given in terms of the form factors
obtained above, as follows
\begin{eqnarray}
\Gamma(t\rightarrow cZ)&=&{1\over 32 \pi m_t^3}(m_t^2-m_Z^2)^2
\Bigl[(2+{m_t^2\over m_Z^2})(F_{Z1}^2+F_{Z1}^{\prime 2}) \nonumber\\
&&-6m_t\,(F_{Z1}F_{Z2}+
F_{Z1}^\prime F_{Z2}^\prime) +(2m_t^2+
m_Z^2)(F_{Z2}^2+F_{Z2}^{\prime 2})\Bigr]\, ,
\label{eq:GammaZ}\\
\Gamma(t\rightarrow c\gamma)&=&{m_t\over 32 \pi }
\Bigl[(2(F_{\gamma1}^2+F_{\gamma1}^{\prime 2}) \nonumber\\
&&-6m_t\,(F_{\gamma1}F_{\gamma2}+
F_{\gamma1}^\prime F_{\gamma2}^\prime) +2m_t^2
(F_{\gamma2}^2+F_{\gamma2}^{\prime 2})\Bigr]\, ,
\label{eq:Gammagamma}\\
\Gamma(t\rightarrow cg)&=&{ m_t\over 24\pi }
\Bigl[(2(F_{g1}^2+F_{g1}^{\prime 2}) \nonumber\\
&&-6m_t\,(F_{g1}F_{g2}+
F_{g1}^\prime F_{g2}^\prime) +2m_t^2
(F_{g2}^2+F_{g2}^{\prime 2})\Bigr]\, .
\label{eq:Gammag}
\end{eqnarray}
In these expressions each form factor receives contributions from both
chargino-squark and neutralino-squark form factors. It is not hard to verify
that for $m_c=0$, the chargino contributions to $F_{1,2}^\prime$ vanish.
(In Ref.~\cite{LOY} there is a factor of $\pi$ missing in the expression for
$\Gamma(t\to cZ)$ and a factor of $1\over3$ missing in $\Gamma(t\to cg)$.)

\section{Numerical Results}
Before we attempt to evaluate the rather lengthy expressions given above, we
would like to consider qualitatively the possibility of dynamical enhancements
of the loop amplitudes. Experience with similar diagrams contributing to the
self-energy of the top quark in supersymmetric theories \cite{KLNR} indicates
possible large corrections when the mass of the top quark equals the sum of
the masses of the other particles leaving the vertex involving the top
quark.\footnote{The sign of
these corrections depends on the observable being calculated: in the case of
$B(t\to cV)$ they are positive, whereas in one-loop supersymmetric corrections
to $\sigma(p\bar p\to t\bar t)$ \cite{KLNR} they were negative.} In the
present case we have vertices with top quarks and: (i) gluinos
and top-squarks in the case of QCD-like contributions (calculated in
Refs.~\cite{LOY,CHK});
(ii) charginos and down-like squarks in the case of `charged' electroweak-like
corrections; and (iii) neutralinos and up-like squarks in the case of `neutral'
electroweak-like corrections. Given the presently-known lower bounds on
the squark (excluding $\tilde t$) and gluino masses ({\em i.e.}, $m_{\tilde
q},m_{\tilde g}>175\,{\rm GeV}$; $m_{\tilde q}\approx m_{\tilde g}>230\,{\rm
GeV}$ \cite{TeVsqg}), this type of enhancement might only be present in the
third type of contribution when $m_t\approx m_\chi+m_{\tilde t_1}$, which
requires a light top-squark whose mass is constrained experimentally to
$m_{\tilde t_1}\gsim60\,{\rm GeV}$ \cite{stop}. The `neutral' electroweak-like
contributions might also be enhanced by large GIM-violating
top-squark--charm-squark mass splittings.  This latter enhancement is also
present in the `charged' electroweak-like contributions.
However, the
`charged' contributions fall short of the `neutral'
electroweak-like corrections for the gluon and photon cases.
Interestingly, the `charged' contribution for the $Z$ is higher than in
the `neutral' case. We first address the neutral
electroweak-like contributions and comment on the charged contributions
afterwards.

The neutral electroweak-like contributions might be enhanced as discussed
above, but this is subject to other mixing factors in the $A_n-F_n$ coupling
functions in Eqs.~(\ref{eq:An})--(\ref{eq:Fgn}) being unsuppressed. At the
root of this question is whether the quark-squark-neutralino couplings might
be flavor non-diagonal as a result of their evolution from the unification
scale down to the electroweak scale. This question might be explored by
considering the squared squark mass matrices at the electroweak scale that
are obtained by renormalization group evolution of a universal scalar
mass at the unification scale~\cite{GH,Duncan}:
\begin{eqnarray}
\widetilde X^2_{iR}&=&M^2_W\,\mu^{(0)}_{iR} I + \mu^{(1)}_{iR} X_i X^\dagger_i
\qquad (i=1,2)
\label{eq:XiR}\\
\widetilde X^2_{1L}&=&M^2_W\,\mu^{(0)}_{1L} I + \mu^{(1)}_{1L} X_1 X^\dagger_1
 + \mu^{(2)}_{1L} X_2 X^\dagger_2
\label{eq:X1L}\\
\widetilde X^2_{2L}&=&M^2_W\,\mu^{(0)}_{2L} I + \mu^{(1)}_{2L} X_1 X^\dagger_1
 + \mu^{(2)}_{2L} X_2 X^\dagger_2
\label{eq:X2L}
\end{eqnarray}
where $i=1$ ($i=2$) corresponds to up-type (down-type) flavors, the
$\mu^{(0,1,2)}$ are RGE-dependent coefficients, and $X_1$ ($X_2$) are the
up-type (down-type) Yukawa matrices. The matrices $B_i\equiv\widetilde U^*_i
U^T_i$, appearing in the equations in Sec.~\ref{sec:formulas} above, are
obtained from the $\widetilde U_i$ matrices that diagonalize $\widetilde
X^2_{iR}$, and the $U_i$ matrices that diagonalize the right-handed quark
mass matrices. Because of the simple form for $\widetilde X^2_{iR}$ in
Eq.~(\ref{eq:XiR}), it can be shown that $\widetilde U_i=U_i$ and
therefore $B_i=I$ \cite{GH}.  [Note that the quark mixing 
matrices $U_i$ and $V_i$
mentioned in this section are different from the chargino mixing
matrices $U_{ij}$ and $V_{ij}$ mentioned in Section 2.] 

The other relevant set of matrices are $\Gamma_i\equiv \widetilde V_i
V^\dagger_i$, obtained from the $\widetilde V_i$ matrices that diagonalize
$\widetilde X^2_{iL}$ and the $V_i$ matrices that diagonalize the left-handed
quark mass matrices. In the case of $\lambda_t\gg\lambda_b$, which requires
$\tan\beta\sim1$, the $X_2 X^\dagger_2\propto\lambda^2_b$ term in
Eqs.~(\ref{eq:X1L},\ref{eq:X2L}) is small compared to the $X_1
X^\dagger_1\propto\lambda^2_t$ term, and therefore the former may be neglected.
This implies that both $\widetilde X^2_{1L}$ and $\widetilde X^2_{2L}$ are
diagonalized by the same matrix: $\widetilde V_1=\widetilde V_2=V_1$, and
therefore $\Gamma_1=\widetilde V_1 V^\dagger_1=I$, whereas $\Gamma_2=\widetilde
V_2 V^\dagger_2=V_1 V^\dagger_2=K$ reduces to the regular CKM matrix \cite{GH}.
As the quark-squark-neutralino couplings in flavor space are proportional to
$\Gamma_i$, we see that for $\lambda_t\gg\lambda_b$ there are no flavor
off-diagonal couplings in the up-quark sector, as required for a unsuppressed
contribution to the `neutral' electroweak-like contributions to $t\to cV$.

One might consider instead a scenario where $\lambda_t\sim\lambda_b$, as would
be consistent with $\tan\beta\gg1$. In this case the $X_2
X^\dagger_2\propto\lambda^2_b$ term in Eqs.~(\ref{eq:X1L},\ref{eq:X2L}) is
no longer negligible and $\Gamma_1\not= I$ is expected. The precise form of
$\Gamma_1$ requires a complicated calculation, essentially solving the matrix
renormalization group equations. For our purposes here it suffices to consider
the following effective form
\begin{equation}
\Gamma_1=\left(
\begin{array}{ccc} 1&0&0\\0&1&\epsilon\\0&-\epsilon&1\end{array}
\right)\ ,
\label{eq:Gamma1}
\end{equation}
where $\epsilon$ parametrizes the size of the ratio $\lambda_b/\lambda_t$.
For moderate values of $\tan\beta$ this form should be adequate ({\em i.e.}
$\epsilon$ not too close to 1). We still expect $\Gamma_2\approx K$.
We assume that the lower $(2\times2)$ right corner of $V_1$ is approximately the
identity to relate $\Gamma_{QL}$ to $\Gamma_1$.

The above forms for $\Gamma_{1,2}$ plus the result $B_{1,2}=I$ above, allow us
to evaluate numerically the branching ratios of Sec.~\ref{sec:formulas}.
Perhaps the most optimistic top factory being contemplated at the moment is a
high-luminosity upgrade of the Tevatron, where studies show that one might be
sensitive to
$B(t\to c\gamma)\approx4\times10^{-4}\,(8\times10^{-5})$~\cite{TeV33},
$B(t\to cZ)\approx 4\times10^{-3}\,(6\times10^{-4})$~\cite{TeV33}, and
$B(t\to cg)\approx5\times10^{-3}\,(1\times10^{-3})$~\cite{tcg} with an
integrated luminosity of $10\,(100)\,{\rm fb}^{-1}$,
where the branching ratios are with respect to $\Gamma(t\to bW)$. These
expected sensitivities will not allow direct tests of the Standard Model
predictions for these processes: $B(t\to c\gamma)^{\rm SM}\sim 10^{-12}$,
$B(t\to cZ)^{\rm SM}\sim10^{-12}$, and $B(t\to cg)^{\rm SM}\sim10^{-10}$
\cite{tcV-SM}, but might uncover virtual new physics effects that enhance
these rates over Standard Model expectations.

Indeed, we generally find that $B(t\to cV)$ greatly exceeds the corresponding
Standard Model contribution, but unfortunately falls below the expected
experimental sensitivities, as was observed also in previous studies of the
QCD-like corrections \cite{CHK}. Specifically, concentrating
on the `neutral' electroweak-like corrections, we have as the dominant inputs
the masses of the charm-squark and top-squark, the top mixing angle,
the mass of the neutralino(s),
and the neutralino composition. The results for $V=g, \gamma$
scale with $\epsilon^2$ as defined
in Eq.~(\ref{eq:Gamma1}); we take $\epsilon=0.5$ for concreteness.
(For $V=Z$, the cross section increases montonically with $\epsilon$ for
$\epsilon\leq0.5$.)
Numerically\footnote{We used the software package {\tt ff}~\cite{vanOld} to
evaluate the Passarino-Veltman functions.}
we find that when $m_t\approx m_\chi+m_{\tilde t_1}$, the branching
ratios are enhanced compared to off-resonance values by a factor of 2-10.
This factor depends on the specific combination of neutralino and
top-squark masses that satisfy this relation (all other parameters being kept
fixed); the enhancement decreases with increasing stop mass and
so is maximised when $m_{\tilde t_1}$ is 
60 {\rm GeV}. 
The off-resonance values themselves are larger than the Standard Model
predictions for not-too-heavy sparticles. We also verify that large $m_{\tilde
c}-m_{\tilde t_1}$ mass splitting enhances the results, because of its
GIM-violating effect. (A useful test of our code is that the branching ratios
go to zero due to the GIM mechanism, if we set squark masses equal.)
With regards to neutralino composition,
the largest branching ratios are obtained for neutralinos with comparable
bino and higgsino admixtures. Increasing the scale of the sparticle masses
typically leads to a rapid decrease in the branching ratios for $g$ and
$\gamma$.
For the $Z$ the decrease is very gradual, as has also
been noted in Ref. \cite{CHK}.
For $g$ and $\gamma$ the cross-section seems
to be maximised for stop mixing angles close to 0 or $\pi$.  
The effects of mixing for $Z$ are very dependent on the
other parameters chosen, such as the neutralino composition, etc.
The effect of varying $\tan\beta$ is of $O(1)$.	 

In varying the
different parameters above we have worked in the most general framework
of the MSSM, in which the various parameters can be varied independently.
In a more specific model, such as one with universal
scalar masses and radiative electroweak symmetry breaking, these
parameters are not all independent.  Although our choice of mixing
matrices was motivated by certain specific scenarios we vary our
parameters freely so as to look for the maximal supersymmetric
contributions.  Furthermore, we choose $m_{\tilde c_{1,2}}$ and 
$m_{\tilde t_2}$ to be $\sim 1 {\rm TeV}$.

For the most optimistic values of the parameters, {\em i.e.}, when the above
enhancing circumstances all simultaneously occur, we find
$B(t\to c\gamma)\lsim 2\times 10^{-7}, B(t\to cZ)\lsim 4\times10^{-7}$ and
$B(t\to cg)\lsim 3\times10^{-5}$. We see that $B(t\to cg)$ is the one
closest to the level of experimental sensitivity expected at the Tevatron,
so perhaps it would be the mode to be first observed at a future
sufficiently sensitive machine. This hope is further enlarged by recalling
that $B(t\to cg)$ receives comparable contributions from the QCD-like
supersymmetric corrections \cite{CHK}, which we have not evaluated here.
Eventually, such a process can be a possible test for supersymmetry.

We have also evaluated the charged electroweak-like
corrections and have found them to be smaller (typically by a factor of
10 or  more) than the neutral electroweak-like corrections discussed
above for $g$ and $\gamma$, but a factor of 10 higher for the $Z$. 
(Again we assume that the lower $(2\times2)$
right corner of $V_2$ is approximately the
identity to relate $\Gamma_{QL}$ to $\Gamma_2$.)
In the case of universal squark masses at the unification scale,
GIM-violating bottom-squark strange-squark mass
differences are generated by RGE evolution, resulting in
shifts to the left-handed down-like squark mass matrices.  The
dominant term is from the second term of Eq.~(\ref{eq:X2L}) which may
be rewritten as
$-|c| K^\dagger (\widehat m_u)^2 K$,
where $\widehat m_u=\{m_u,m_c,m_t\}$ and $|c|\leq1$ is an RGE-dependent
constant.  Inserting the values of the CKM matrix elements we find
(approximately):
$m^2_{\tilde s_L}\to m^2_{\tilde s_L}-|c|(m_t/5)^2$ and  $m^2_{\tilde
b_L}\to m^2_{\tilde b_L}-|c|m_t^2$. 
Choosing the maximal ($|c|=1$) mass splittings, we find
$B(t\to c\gamma)\lsim10^{-8}$,
$B(t\to cZ)\lsim2\times10^{-6}$, and
$B(t\to cg)\lsim10^{-7}$ for
squark masses as low as
experimentally allowed.  (These results are not much altered even if one drops
the assumption of universal squark masses at the unification scale.)
The numerical results for the
charged electroweak-like corrections cannot be compared with the
corresponding ones in Ref.~\cite{LOY} because, as we explained above,
the formulas presented in Ref.~\cite{LOY} are inconsistent with gauge
invariance constraints.

We finally try to connect up with the recent literature in Ref.~\cite{tcg},
where in addition to the $t\to cg$ decay mode, people have considered hadronic
processes like $p\bar p\to t\bar c$, which might be more easily detectable.
These works ignore any substructure that the $t$-$c$-$g$ vertex might have,
and replace it all by an effective scale $\Lambda$, defined for instance by
the relation $\Gamma(t\to cg)=8\alpha_s m^3_t/3\Lambda^2$. A given branching
ratio obtained in the supersymmetric theory then corresponds to a scale
$\Lambda$ in the effective theory. Dividing this expression by $\Gamma(t\to
bW)$ we find that $\Lambda\approx1\,{\rm TeV}/\sqrt{B(t\to cg)}$. Therefore
a supersymmetric prediction of $B(t\to cg)\sim10^{-5}$ corresponds to
$\Lambda\sim300\,{\rm TeV}$. The point of this exercise is to note how
misleading such estimates of new-physics scales might be, as this actually
corresponds in our case to sparticle masses of a few hundred GeV!

\section*{Acknowledgments}
The work of R.R. has been supported by the World Laboratory. The work of J.L.
has been supported in part by DOE grant DE-FG05-93-ER-40717 and that of D.V.N.
by DOE grant DE-FG05-91-ER-40633.  We would also like to thank the referee
for pointing out the possible relevance of the `charged' contributions.

\clearpage

\end{document}